\documentclass[pra,twocolumn,twoside,nopacs,nokeys,superscriptaddress]{revtex4}

\usepackage{graphicx,epic,eepic,epsfig,amsmath,latexsym,amssymb,verbatim}

\usepackage{theorem}
\newtheorem{definition}{Definition}[section]
\newtheorem{proposition}[definition]{Proposition}
\newtheorem{lemma}[definition]{Lemma}

\newtheorem{theorem}[definition]{Theorem}
\newtheorem{corollary}[definition]{Corollary}
\newtheorem{conjecture}[definition]{Conjecture}

\newtheorem{rem}[definition]{Remark}

\def\squareforqed{\hbox{\rlap{$\sqcap$}$\sqcup$}}
\def\qed{\ifmmode\squareforqed\else{\unskip\nobreak\hfil
\penalty50\hskip1em\null\nobreak\hfil\squareforqed
\parfillskip=0pt\finalhyphendemerits=0\endgraf}\fi}
\def\endenv{\ifmmode\;\else{\unskip\nobreak\hfil
\penalty50\hskip1em\null\nobreak\hfil\;
\parfillskip=0pt\finalhyphendemerits=0\endgraf}\fi}
\newenvironment{proof}{\noindent \textbf{{Proof~} }}{\qed}

\mathchardef\ordinarycolon\mathcode`\:
\mathcode`\:=\string"8000
\def\vcentcolon{\mathrel{\mathop\ordinarycolon}}
\begingroup \catcode`\:=\active
  \lowercase{\endgroup
  \let :\vcentcolon
  }

\newcommand{\nc}{\newcommand}
\nc{\rnc}{\renewcommand}
\nc{\beq}{\begin{equation}}
\nc{\eeq}{{\end{equation}}}
\nc{\beqa}{\begin{eqnarray}}
\nc{\eeqa}{\end{eqnarray}}
\nc{\lbar}[1]{\overline{#1}}
\nc{\bra}[1]{\langle#1|}
\nc{\ket}[1]{|#1\rangle}
\nc{\ketbra}[2]{|#1\rangle\!\langle#2|}
\nc{\braket}[2]{\langle#1|#2\rangle}
\nc{\proj}[1]{| #1\rangle\!\langle #1 |}
\nc{\avg}[1]{\langle#1\rangle}
\rnc{\max}{\operatorname{max}}
\nc{\Rank}{\operatorname{Rank}}
\nc{\smfrac}[2]{\mbox{$\frac{#1}{#2}$}}
\nc{\tr}{\operatorname{Tr}}
\nc{\ox}{\otimes}
\nc{\dg}{\dagger}
\nc{\dn}{\downarrow}
\nc{\cA}{{\cal A}}
\nc{\cB}{{\cal B}}
\nc{\cC}{{\cal C}}
\nc{\cD}{{\cal D}}
\nc{\cE}{{\cal E}}
\nc{\cF}{{\cal F}}
\nc{\cG}{{\cal G}}
\nc{\cH}{{\cal H}}
\nc{\cI}{{\cal I}}
\nc{\cJ}{{\cal J}}
\nc{\cK}{{\cal K}}
\nc{\cL}{{\cal L}}
\nc{\cM}{{\cal M}}
\nc{\cN}{{\cal N}}
\nc{\cO}{{\cal O}}
\nc{\cP}{{\cal P}}
\nc{\cR}{{\cal R}}
\nc{\cS}{{\cal S}}
\nc{\cT}{{\cal T}}
\nc{\rU}{{\cal U}}
\nc{\cX}{{\cal X}}
\nc{\cZ}{{\cal Z}}
\nc{\csupp}{{\operatorname{csupp}}}
\nc{\qsupp}{{\operatorname{qsupp}}}
\nc{\var}{\operatorname{var}}
\nc{\rar}{\rightarrow}
\nc{\lrar}{\longrightarrow}
\nc{\polylog}{\operatorname{polylog}}
\nc{\1}{\openone}

\def\d{\delta}
\def\e{\epsilon}

\nc{\RR}{{{\mathbb R}}}
\nc{\CC}{{{\mathbb C}}}
\nc{\FF}{{{\mathbb F}}}
\nc{\NN}{{{\mathbb N}}}
\nc{\ZZ}{{{\mathbb Z}}}
\nc{\PP}{{{\mathbb P}}}
\nc{\QQ}{{{\mathbb Q}}}
\nc{\UU}{{{\mathbb U}}}
\nc{\WW}{{{\mathbb W}}}
\nc{\EE}{{{\mathbb E}}}
\nc{\II}{{{\mathbb I}}}
\nc{\id}{{\operatorname{id}}}

\nc{\ob}[1]{#1}

\begin{document}

\title{On the quantum, classical and total\protect\\
       amount of correlations in a quantum state}

\author{Berry Groisman}\email{b.groisman@bris.ac.uk}
\affiliation{H. H. Wills Physics Laboratory, Royal Fort, Tyndall Avenue, Bristol BS8 1TL, U.K.}

\author{Sandu Popescu}\email{s.popescu@bris.ac.uk}
\affiliation{H. H. Wills Physics Laboratory, Royal Fort, Tyndall Avenue, Bristol BS8 1TL, U.K.}
\affiliation{Hewlett-Packard Laboratories, Stoke Gifford, Bristol BS12 6QZ, U.K.}

\author{Andreas Winter}\email{a.j.winter@bris.ac.uk}
\affiliation{Department of Mathematics, University of Bristol, Bristol BS8 1TW, U.K.}

\date{31st January 2005}
%begun: 29 June 2004

\begin{abstract}
  We give an operational definition of the quantum, classical and total amount
  of correlations in a bipartite quantum state.
  We argue that these quantities can be defined
  via the amount of work (noise) that is required to erase (destroy)
  the correlations:
  for the total correlation, we have to erase completely, for the
  quantum correlation one has to erase until a separable state is obtained,
  and the classical correlation is the maximal correlation
  left after erasing the quantum correlations.
  \par
  In particular, we show that the total amount of correlations is equal to
  the quantum mutual information, thus providing it with a direct operational
  interpretation for the first time. As a byproduct, we obtain a direct,
  operational and elementary proof of strong subadditivity of quantum entropy.
\end{abstract}

\maketitle

\section{Introduction}
\label{sec:intro}
Landauer~\cite{Landauer}, in analysing the
physical nature of (classical) information,
showed that the amount of information stored, say, in a computer's memory, is
proportional to the work required to erase the memory (reset to zero all the bits).
These ideas were further developed by other researchers
(most prominently Bennett)
into a deep connection of classical information and thermodynamics
(see~\cite{Bennett} for a recent survey).
Here we follow Landauer's idea in analysing quantum information:
we want to measure correlation by the (thermodynamical) effort
required to erase (destroy) it.

The main idea of our paper can be understood on a simple example. Consider a
maximally entangled state of two qubits (equivalent to a singlet)
\begin{equation}
  \ket{\Phi^+} = \frac{1}{\sqrt2}(\ket{0}_A\ox\ket{0}_B+\ket{1}_A\ox\ket{1}_B).
  \label{eq:epr}
\end{equation}

Usually this state is seen as containing $1$ ebit, i.e., one bit of entanglement,
based on the asymptotic theory of pure state entanglement~\cite{BBPS}.
The temptation is to think that it contains $1$ bit of correlation, and that
this correlation is in pure quantum form
(which can be used either quantumly
--- e.g., for teleportation --- or
to obtain one perfectly correlated classical bit).

We will argue however that this state
contains in fact 2 bits of correlation --- 1 bit of entanglement and 1 bit of
remaining (secret) classical correlations, as follows.

Suppose that Alice wants to erase the entanglement between her bit and Bob's.
She can do this by applying $1$ bit of \emph{randomness}:
she applies to her qubit one of two unitary transformations $\1$ or $\sigma_z$
with equal probability. By this the pure state in eq.~(\ref{eq:epr})
becomes a mixture

$$\rho = \frac{1}{2} \proj{\Phi^+} + \frac{1}{2} \proj{\Phi^-},$$
where

$$\ket{\Phi^-}=\frac{1}{\sqrt{2}} (\ket{0}_A\ox\ket{0}_B - \ket{1}_A\ox\ket{1}_B).$$

This mixed state is disentangled because it is identical with a mixture
of two direct product states

$$\rho=\frac{1}{2} \proj{0}_A\ox\proj{0}_B + \frac{1}{2} \proj{1}_A\ox\proj{1}_B.$$
But although the entanglement is now gone, Alice and Bob's qubits are still correlated.
Indeed, $\rho$ contains now 1 bit of purely classical correlations; furthermore,
these correlations are secret since they are not correlated with any third party,
such as an eavesdropper.

To also erase these classical correlations Alice has to ``work'' more.
She can do this by randomly applying a ``bit flip'' to $\rho$,
that is, applying at random,
with equal probability either $\1$ or $\sigma_x$. This brings the
state to
$$\rho' = \frac{1}{2}\1_A \otimes \frac{1}{2}\1_B,$$
where qubit A is completely independent from qubit B.

To summarise, two bits of erasure (or, depending on the point of view,
``bits of noise'', or ``error''), are required to completely erase the
correlations in the singlet. The first bit erases the entanglement and
the second erases the classical secret correlations. We then say that the
singlet contains 1 bit of pure entanglement, \emph{and} 1 bit of secret
classical correlations. The \emph{total} amount of correlations is 2 bits.

A couple of remarks concerning the connection to Landauer's theory of information
erasure: just as Landauer for information (entropy!), our approach quantifies
correlations via their \emph{robustness} against destruction. However, there
seems to be a contradiction: whereas Landauer considers \emph{resetting}
the memory to a standard state (and we take for granted that one can
generalise his argument to quantum memory), effectively exporting
--- ``dissipating'' --- the entropy of the system, we \emph{inject}
entropy into it.
This is actually only an apparent contradiction, as can be seen easily once
we realise that in the above example we tacitly assumed that Alice \emph{forgets}
which Pauli operator she has applied. Indeed, we can present what she
does in more detail as follows: she has a reservoir
of random bits, which she uses to apply one of the Pauli operators as above
in a reversible way (by a quantum-controlled unitary). This step does not
affect the correlations between $A$ and $B$. Only when she decides to erase
(forget) the random bits, the correlations are affected, as we have shown
above. Now it is evident that the entropy pumped into the
state is equal to the Landauer erasure cost of the random bits.

In this paper we develop these ideas, as follows.

For an arbitrary bipartite quantum state $\rho_{AB}$ the quantum mutual
information is defined as
$$I(A:B)=S(\rho_A) + S(\rho_B) - S(\rho_{AB}).$$
(The name is taken from Cerf and Adami~\cite{Cerf:Adami}, but
Stratonovich~\cite{stratonovich} has considered this quantity
already in the mid-60s.)

While this definition is formally very simple, an operational
interpretation for it was hitherto missing \cite{CMI} (at least
not for the quantity itself with given state; it plays however a
crucial role in the formula for the entanglement-assisted capacity
of a quantum channel~\cite{BSST}). We show here that the total
amount of correlations, as measured by the minimal rate of
randomness that are required to completely erase all the
correlations in $\rho_{AB}$ (in a many-copy scenario), is equal to
the quantum mutual information. This is the main result of
section~\ref{sec:bi-total}. As an important consequence of this
result we shall demonstrate that it leads to the {\it strong
subadditivity} of von Neumann entropy.

In our above example this amount of total correlations divides
neatly into the amount required to obliterate the quantum
correlations ($1$ bit), and the amount to take the resulting
separable state to a product state ($1$ bit). We will follow on this
in our discussion contained in section~\ref{sec:bi-entanglement},
where we use this approach to define the quantum and the
classical correlations in a state, and conjecture how they compare
with the total correlations.

Section~\ref{sec:comparison} contains some observations how
the total, quantum and classical correlation as defined here
relates to other such measures.

Then, in section~\ref{sec:multiparty}, we extend our considerations
to correlations (quantum and classical) of more than two players,
after which we conclude.

An appendix quotes the technical results about typical subspaces
and our main tool, an operator version of the classical
Chernoff bound, which are used repeatedly, as well
as miscellaneous proofs.

\section{Total bipartite correlations}
\label{sec:bi-total}
As explained in the introduction, we want to add randomness to a state
$\rho=\rho_{AB}$ of a bipartite system $AB$ (with local Hilbert space
dimensions $d_A,\,d_B<\infty$)
in such a way as to make it into a product state.
In fact, we shall consider $n\rar\infty$ many
copies of $\rho$, and be content with achieving decorrelation
(product state) approximately (but arbitrarily good in the asymptotic
limit).

In detail, the randomisation will be engineered by an ensemble of
local unitaries, $\{p_i,U_i\ox V_i\}_{i=1}^N$, to which is
associated the randomising map
\begin{equation}
  \label{eq:R}
  R:\tau \longmapsto \sum_{i=1}^N p_i (U_i\ox V_i)\tau(U_i\ox V_i)^\dg.
\end{equation}
We call the class of such completely
positive and trace preserving (cptp) maps on $AB$ ``coordinated
local unitary randomising'' (COLUR).
Considering that our object is to study the
correlation between $A$ and $B$, it may seem a bit suspicious to
allow coordinated application of $U_i$ and $V_i$ at the two sites.
Hence we define $A$-LUR to be those maps where all $V_i=\1$, and
$B$-LUR those where all $U_i=\1$ --- because they can be
implemented by application of noise strictly locally at $A$ or $B$
alone, respectively. The combination of an $A$-LUR with a $B$-LUR
map (i.e., independent local noise at either side) we call
simply ``local unitary randomising'' (LUR).

We say that $R$ \emph{$\e$-decorrelates} a state $\rho$ if there
is a product state $\omega_A \ox \omega_B$ such that
\begin{equation}
  \label{e-decorrelation}
  \bigl\| R(\rho) - \omega_A\ox\omega_B \bigr\|_1 \leq \e,
\end{equation}
where $\|\cdot\|_1$ is the trace norm of an operator,
i.e.~the sum of the absolute values of the eigenvalues. For
technical reasons, when we study the asymptotics of such
transformations (i.e., acting on $n$ copies of the state $\rho$),
we will demand that the \emph{output} of the map $R$ (and similar
maps studied below) is supported on a space of dimension $d^n$,
for all $n$, with some finite $d$.

How to account for the amount of noise introduced: from the point
of view of the ensemble of unitaries, the most conservative option
will be to take $\log N$, the space required to identify the
element $i$ uniquely. A smaller, and in the many-copy asymptotic
meaningful, quantity would be $H(p) = -\sum_i p_i\log p_i$. Note
however that they are not uniquely associated with the randomising
map $R$. However, Schumacher~\cite{S-exchange}, and earlier
Lindblad~\cite{Lindblad}, have proposed a measure of the entropy
of a cptp map $T$ injects into the system $P$ on which it acts:
for this purpose, one has to introduce an \emph{environment} $E$,
which is initially in a pure state, and to fix a \emph{reference
system} $Z$, which purifies $\rho_P$ to $\ket{\psi}_{ZP}$ --- note
that all such purifications are related via unitaries on $Z$.
Then, the \emph{entropy exchange} is defined as
$$S_e(T,\rho_P) := S\bigl( (\id_Z\ox T_P)\proj{\psi} \bigr).$$
It is the entropy the \emph{environment} (initially in a pure state)
acquires in a unitary dilation of the cptp map. In
this paper, $P$ will be a bipartite system $AB$.

Based on elementary properties of the von Neumann entropy, one can
see that for every randomising map $R$ as above, and every state $\rho$,
\begin{equation}
  \label{3measures}
  \log N \geq H(p) \geq S_e(R,\rho).
\end{equation}

\begin{proposition}
  \label{prop:bi-total:lower}
  Consider any COLUR map on the bipartite system $A^nB^n$,
  $$R:\tau \longmapsto \sum_{i=1}^N p_i (U_i\ox V_i)\tau(U_i\ox V_i)^\dg,$$
  which $\e$-decorrelates $\rho^{\ox n}$. Then the entropy exchange
  of $R$ relative to $\rho^{\ox n}$ is lower bounded
  \begin{equation}
    \label{eq:entropy-exchange:lower}
    S_e\bigl( R,\rho^{\ox n} \bigr)
            \geq n\bigl( I(A:B) - 3\e\log d - \eta(3\e) \bigr),
  \end{equation}
  where
  \begin{equation*}
    \eta(x) := \begin{cases}
                 -x\log x          & \text{for }x\leq \frac{1}{e}, \\
                 \frac{1}{e}\log e & \text{for }x\geq \frac{1}{e}.
               \end{cases}
  \end{equation*}
  In particular, the right hand side is also a lower bound
  on $H(p)$, and even more so on $\log N$.
\end{proposition}
\begin{proof}
  First of all, because $R$ acts locally,
  \begin{equation*}
    R_A := \tr_B R(\rho^{\ox n})
         = \sum_{i=1}^N p_i U_i \rho_A^{\ox n} U_i^\dg,
  \end{equation*}
  and similarly for $R_B := \tr_A R(\rho^{\ox n})$. Hence we have
  (using the concavity of the von Neumann entropy)
  \begin{equation}\label{localineq}
    S(R_A) \geq n S(\rho_A),
    \quad
    S(R_B) \geq n S(\rho_B),
  \end{equation}
  On the other hand, we can argue that $R(\rho^{\ox n})$ is very close
  to $R_A \ox R_B$. Indeed, from eq.~(\ref{e-decorrelation}) it follows that
  \begin{equation*}
    \bigl\| R_A - \omega_A \bigr\|_1
       \leq \bigl\| R(\rho^{\ox n}) - \omega_A\ox\omega_B \bigr\|_1 \leq \e.
  \end{equation*}
  Similarly,
  \begin{equation*}
    \bigl\| R_B - \omega_B \bigr\|_1 \leq \e.
  \end{equation*}
  Thus, by the triangle inequality,
  $$\bigl\| R_A\ox R_B - \omega_A\ox\omega_B \bigr\|_1 \leq 2\e,$$
  and we get
  \begin{equation}
    \bigl\| R(\rho^{\ox n}) - R_A\ox R_B \bigr\|_1 \leq 3\e.
  \end{equation}
  Hence, by the Fannes inequality~\cite{Fannes74},
  \begin{equation}
    \label{eq:entropy-lower-fannes}
    S(R_A)+S(R_B)-S\bigl( R(\rho^{\ox n}) \bigr) \leq 3\e\log d^n + \eta(3\e).
  \end{equation}
  Taking into account eq.~(\ref{localineq}) we obtain
  \begin{equation}
    \label{eq:entropy-lower}
    S\bigl( R(\rho^{\ox n}) \bigr)
           \geq n\bigl( S(\rho_A)+S(\rho_B) - 3\e\log d - \eta(3\e) \bigr).
  \end{equation}
  Here we use the fact that multiplying the last term in
  eq.~(\ref{eq:entropy-lower-fannes}) by $n$ will only weaken
  the inequality.
  Now, introduce a purifying reference system $Z$ for our state:
  $\rho = \tr_Z\psi$, with a pure state $\psi=\proj{\psi}$ on $ZAB$.
  Then the randomising map acts on $A^nB^n$, producing the state
  $$\Omega = \bigl(\id_Z^{\ox n}\ox R \bigr)\bigl(\psi^{\ox n}\bigr)$$
  on $Z^nA^nB^n$. So, by definition of the entropy exchange,
  \begin{equation*}\begin{split}
    S_e\bigl( R,\rho^{\ox n} \bigr)
                        &= S(\Omega_{Z^nA^nB^n})                   \\
                        &\geq S(\Omega_{A^nB^n}) - S(\Omega_{Z^n}) \\
                        &=    S\bigl( R(\rho^{\ox n}) \bigr)
                                     - S\bigl( \rho^{\ox n} \bigr) \\
                        &\!\!\!\!\!\!\!\!
                         \geq n\bigl( S(\rho_A)+S(\rho_B)-S(\rho)
                                           - 3\e\log d - \eta(3\e) \bigr),
  \end{split}\end{equation*}
  where in the second line we have used the Araki-Lieb (or
  triangle) inequality~\cite{AL71},
  and in the third line the fact that $R$ acted only on $A^nB^n$,
  i.e. initially $S(\rho_{Z^n})=S(\rho_{A^nB^n})$;
  in the last line we have inserted eq.~(\ref{eq:entropy-lower}).
\end{proof}

On the other hand, we have:
\begin{proposition}
  \label{prop:bi-total:upper}
  For any state $\rho$ and $\e>0$ there exists, for all
  sufficiently large $n$, an $A$-LUR map
  $$R:\tau \longmapsto \frac{1}{N} \sum_{i=1}^N (U_i\ox \1)\tau(U_i\ox \1)^\dg$$
  on $A^nB^n$, which $\e$-decorrelates $\rho^{\ox n}$, and with
  $$\log N \leq n\bigl( I(A:B) + \e \bigr).$$
\end{proposition}
\begin{proof}
  For large $n$, we change the state $\rho^{\ox n}$
  very little by restricting it to its typical subspace,
  with projector $\Pi$ (see appendix A),
  and even restricting the systems $A^n$ ($B^n$) to the local typical
  subspaces of $\rho_A^{\ox n}$ ($\rho_B^{\ox n}$), with projector
  $\Pi_A$ ($\Pi_B$):
  \begin{equation}
    \label{eq:hat-rho}
    \widehat{\rho} := (\Pi_A\ox\Pi_B)\Pi\rho^{\ox n}\Pi(\Pi_A\ox\Pi_B).
  \end{equation}
  By definition of the typical subspace projectors,
  $$\bigl\| \widehat{\rho} - \rho^{\ox n} \bigr\|_1
                          \leq \e + \sqrt{8\cdot 2\e} \leq 5\sqrt{e},$$
  using the ``gentle measurement lemma''~\ref{lemma:gentle}.

  From the properties of the typical projectors (see again appendix A)
  we obtain that $\widehat{\rho}$ is an operator of trace $\geq 1-3\e$
  supported on a tensor product of (typical sub-) spaces of
  dimensions $D_A \leq 2^{n\bigl( S(\rho_A)+\e \bigr)}$ and
  $D_B \leq 2^{n\bigl( S(\rho_n)+\e \bigr)}$, and such that
  $$\widehat{\rho} \leq \frac{1}{D}\Pi_A\ox\Pi_B,$$
  where $D = 2^{n\bigl( S(\rho)-\e \bigr)}$. It is for this latter
  property that we needed to put the global typical projector
  $\Pi$ in the definiton of $\widehat{\rho}$, eq.~(\ref{eq:hat-rho}).

  For the following argument we will also need a lower bound on
  the reduced state on $B$, which we engineer by a further reduction:
  define the projection $\Pi_B'$ on the subspace where
  $\tr_A\widehat{\rho} \geq \e/D_B$, and let
  $$\widetilde{\rho} := (\1_A\ox\Pi_B')\widehat{\rho}(\1_A\ox\Pi_B').$$
  Then it is immediate that
  $\tr\widetilde{\rho} \geq \tr\widehat\rho - \e \geq 1-4\e$,
  hence by the gentle measurement lemma~\ref{lemma:gentle}
  $$\bigl\| \widetilde\rho - \widehat\rho \bigr\|_1 \leq \sqrt{8\e},$$
  and we can keep for later reference the approximation
  \begin{equation}
    \label{eq:approx}
    \bigl\| \widetilde{\rho} - \rho^{\ox n} \bigr\|_1 \leq 8\sqrt{\e}.
  \end{equation}
  Observe that we have defined all these projections in such a way that
  $$\omega_B' := \tr_A\widetilde\rho \geq \frac{\e}{D_B}\Pi_B'.$$

  Now take any ensemble of unitaries, $\bigl\{ p({\rm d}U),U \bigr\}$, such that
  for all state $\varphi$ from the typical subspace of $\rho_A^{\ox n}$,
  $$\int_U p({\rm d}U) U\varphi U^\dg = \frac{1}{D_A}\Pi_A =: \omega_A$$
  (a \emph{private quantum channel} in the terminology of~\cite{private:quchannel}),
  for example, the discrete Weyl operators on the typical subspace
  of $\rho_A^{\ox n}$, but all unitaries on that subspace with corresponding
  Haar measure are good, as well. (The unitaries can behave in any way outside
  the subspace.)
  By elementary linear algebra,
  $$\int_U p({\rm d}U)(U\ox\1)\widetilde{\rho}(U^\dg\ox\1)
                                      = \omega_A\ox\omega_B'.$$

  Now we show, using the ``operator Chernoff bound'', lemma~\ref{lemma:opChernoff}
  in appendix A, that we can select a small subensemble of these unitaries
  doing the same job to sufficient approximation (this is an argument
  like those used in~\cite{HLSW}). To this end, we understand
  Alice's local unitary $U$ as random variable with distribution $p({\rm d}U)$,
  and define the operator valued random variable

  $$X := D (U\ox\1)\widetilde{\rho}(U^\dg\ox\1).$$

  By the above, $0\leq X\leq \1$ and
  $$\EE X = D \omega_A \ox \omega_B'
         \geq \e 2^{-n\bigl( I(A:B)+3\e \bigr)}\Pi_A\ox\Pi_B'.$$

  Thus, if $X_1,\ldots,X_N$ are independent realisations of $X$,
  lemma~\ref{lemma:opChernoff} yields
  \begin{equation*}\begin{split}
    \Pr&\left\{ \frac{1}{N}\sum_{i=1}^N X_i
                   \not\in \left[ (1-\e)\EE X;(1+\e)\EE X \right] \right\} \\
       &\phantom{==}
        \leq 2 d_A^n d_B^n
              \exp\left( -N {\e 2^{-n\bigl( I(A:B)+3\e \bigr)}\e^2}/{2} \right)
  \end{split}\end{equation*}
  where the factor $2$ on the right hand side follows from adding the two
  probability bounds of lemma \ref{lemma:opChernoff}.
  For $N = 2^{n\bigl( I(A:B)+4\e \bigr)}$ or larger
  (and sufficiently large $n$)
  this is smaller than $1$, and we can conclude that there
  exist $U_1,\ldots,U_N$ from the a priori ensemble such that
  $$(1-\e)\omega_A\ox\omega_B'\!
        \leq \frac{1}{N}\sum_{i=1}^N (U_i\ox\1)\widetilde{\rho}(U_i\ox\1)^\dg
    \leq\! (1+\e)\omega_A\ox\omega_B'.$$
  Note, that it is enough to show that this probability is just smaller than one, i.e.
  that at least one such set of unitaries exists.

  Putting this together with eq.~(\ref{eq:approx}), we get
  $$\left\| \frac{1}{N}\sum_{i=1}^N (U_i\ox\1)\rho^{\ox n}(U_i\ox\1)^\dg
                                              - \omega_A\ox\omega_B' \right\|_1
       \leq \e + 8\sqrt{\e},$$
  hence for the state $\omega_B := \omega_B'/\tr{\omega_B'}$,
  $$\left\| \frac{1}{N}\sum_{i=1}^N (U_i\ox\1)\rho^{\ox n}(U_i\ox\1)^\dg
                                              - \omega_A\ox\omega_B \right\|_1
                                                             \leq 5\e + 8\sqrt{\e}.$$
  The last inequality shows that the map $R$ we have constructed,
  does indeed $(5\epsilon+8\sqrt{\epsilon})$-decorrelate
  $\rho^{\ox n}$.
\end{proof}

\par\medskip
Putting eq.~(\ref{3measures}) and
propositions~\ref{prop:bi-total:lower}
and~\ref{prop:bi-total:upper} together, we obtain the (robust)
asymptotic measure of total correlation in a quantum state:

\begin{theorem}
  \label{thm:bi-total}
  The total correlations in a bipartite state $\rho_{AB}$, as measured
  by the asymptotically minimal amount of local noise one has to
  add to turn it into a product (let us denote this $C_{\rm er}(\rho)$,
  the correlation of erasure of $\rho$),
  is $I(A:B)=S(\rho_A)+S(\rho_B)-S(\rho_{AB})$.
  Mathematically,
  \begin{equation*}\begin{split}
    &\sup_{\e>0}\liminf_{n\rar\infty}\frac{1}{n}
                \min\bigl\{ S_e(R,\rho^{\ox n}) : R\ \e\text{-decorr. COLUR} \bigr\} \\
    &\phantom{==}
     =\sup_{\e>0}\limsup_{n\rar\infty}\frac{1}{n}
                 \min\bigl\{ \log N : R\ \e\text{-decorr. $A$-LUR} \bigr\} \\
    &\phantom{==}
     =I(A:B).
  \end{split}\end{equation*}
  So, whether we allow general LUR ensembles or ones restriced to $A$ (or $B$),
  whether we count conservatively the size of the ensemble, $\log N$, or
  be lax and charge only the entropy exchange, and whether we define
  the best rate optimistically or pessimistically, it all comes
  down to the quantum mutual information as the optimal noise (erasure)
  rate to remove the total correlation.
  \qed
\end{theorem}

In passing we note that this implies the perhaps surprising result
that the three ways of measuring the noise in eq.~(\ref{3measures}),
are asymptotically equivalent, as expressed in
propositions~\ref{prop:bi-total:lower}
and~\ref{prop:bi-total:upper}. In~\cite{oppenheim:reznik} the authors
argue that the entropy exchange is a way of measuring the noise of
a cptp map based on compressibility --- it seems to us that the connection
to that work is the following:
while one can always change the basis of the environment to interpret the
entropy exchage as the ``entropy of Kraus operators acting'', this
change of basis will turn our initially unitary Kraus operators into
something else. We instead want to modify the cptp map so as
to preserve the entropy exchange \emph{and} unitarity of the Kraus
operators.

We now want to present a line of thought intended to reconcile our
earlier doubts whether allowing coordinated LUR would be a well-behaved
concept. This is based on the realisation that providing the players with
the perfectly correlated data $i$ (with probability $p_i$) is effectively
giving them another state $\gamma=\sum_i p_i \proj{i}_A\ox\proj{i}_B$.
This gives us the idea of regarding the situation as a kind of
catalysis; the task, for given (general) $\gamma$,
is to decorrelate $\rho\ox\gamma$, but we will have to
discount the overhead $C_{\rm er}(\gamma)$ of just erasing
the correlations in $\gamma$.

So, we really want to consider the infimum (over all $\gamma$),
of the erasure cost of $\rho\ox\gamma$ minus the cost of $\gamma$.
Of course, in the light of our theorem~\ref{thm:bi-total}, this is
$I(A:B)$ (which means that allowing catalysis does not change
the content of our theorem). Conceptually, however, we gain
an insight: supposing we allow only LUR in the randomisation,
then giving the parties a perfect correlation $\gamma$
allows them the following strategy: they use the perfect
correlation to implement a general COLUR map to erase the correlations
in $\rho$ and after this the one in $\gamma$. We don't need to know
how much the latter costs because we subtract the same cost anyway.

Thus, even though we may be restricted to LUR at first, the
availability of appropriate $\gamma$ in a catalytic scenario
effectively motivates consideration of general COLUR maps.
It is a nice observation, though, that in theorem~\ref{thm:bi-total}
we can locally restrict to $A$-LUR without the need to resort
to catalysts.

%We note that in the asymptotic regime three original cost measures
%are equivalent. The following operational interpretation may be
%given to $S_e$. In addition to the system $Z$ we add some {\it
%environment} system $E$ which is initially in a pure state
%$\ket{0}_E$. Then, the randomisation $R$ on $AB$ is interpreted
%as a noise induced by the unitary acting on $ABZE$. This unitary
%transforms $\ket{0}_E$ into a mixture, which on the
%asymptotic limit is $\sum_i p_i \proj{i}_{E}$. Thus, $S_e$ is the
%entropy gain of the environment as a result of the noise induced.

\begin{rem}
  \label{rem:LUN}
  {\rm
  It may be worth noting that our lower bound in
  proposition~\ref{prop:bi-total:lower} is valid for an even
  larger class of operations, namely ``local unital'' (LUN) cptp
  maps: these are compositions of unital (i.e., identity preserving)
  maps locally at $A$ and at $B$. This is because all we need
  for the argument is that the local entropies of Alice and Bob
  can only increase under the map, which is exactly the property
  of unital cptp maps;  %% REFERENCE???
  the rest of the proof is the same
  (observe in particular that entropy exchange makes sense for
  whatever cptp map we have, not just mixtures of unitaries!).
  Cleary LUR is a subset of LUN, and we can even emulate COLUR
  maps by including catalysis in the sense of the previous
  remarks.
  \par
  We can interpret this result intuitively using our explanation of our
  approach in terms of (reversible) local unitaries and Landauer
  erasure, as given in the introduction.
  Namely, it is well-known that unital maps $T$ are exactly those
  which admit a dilation
  $$T(\varphi) = \tr_E\left(
                        U\left( \varphi\ox\frac{1}{d_E}\1_E \right)U^\dg
                       \right).$$
  Hence, the local unital maps of Alice and Bob can be understood as
  reversibly interacting their registers with local noise, and
  subsequent erasure of that noise. The cost of the latter is bounded
  by the entropy exchange.
  }
\end{rem}

\begin{corollary}[Strong subadditivity]
  \label{cor:SSA}
  For any tripartite state $\rho_{ABC}$,
  \begin{equation*}\begin{split}
    I(A:C|B) &= S(\rho_{AB}) + S(\rho_{BC}) \\
             &\phantom{=}
                - S(\rho_{ABC}) - S(\rho_B) \geq 0.
  \end{split}\end{equation*}
\end{corollary}
\begin{proof}
  The strong subadditivity inequality as expressed above is
  equivalent to
  $$I(A:BC) \geq I(A:B).$$
  However, by theorem~\ref{thm:bi-total} above, the left hand side
  is the minimum local noise necessary and sufficient to
  asymptotically decorrelate $A$ from $BC$, and we may consider
  an $A$-LUR for this, i.e., randomisation acting only on $A$.
  Since a map which $\e$-decorrelates $\rho_{A|BC}$ surely
  also $\e$-decorrelates $\rho_{AB}$, this minimum noise
  is larger or equal than the minimum noise to decorrelate
  the latter state, which is the right hand side, once
  more by theorem~\ref{thm:bi-total}.
  \par
  Observe that the proof of theorem~\ref{thm:bi-total}
  did not invoke strong subadditivity: in the lower bound,
  proposition~\ref{prop:bi-total:lower}, we have only used
  concavity (Schur convexity) and subadditivity of the
  entropy; in the upper bound, proposition~\ref{prop:bi-total:upper},
  only typical subspaces and random coding were employed.
\end{proof}

\begin{rem}
  \label{rem:total:LOPC-mono}
  {\rm
  While it is worth noting that in our noise model we have not allowed
  communication between the parties, and that indeed (and unsurprisingly)
  communication can decrease as well as increase the total correlation,
  our result shows that the total correlation $C_{\rm}(\rho)$ is
  indeed monotonic under local operations and \emph{public} communication
  (LOPC) in the following sense.

  Every LOPC is a succession of steps of the form that Alice (Bob)
  performs a quantum instrument~\cite{davies:lewis} locally,
  transforming the state $\rho$ into an ensemble $\bigl\{p_i,\rho_i\bigr\}$,
  of which she (he) communicates $i$ to the other party. In
  general, such local quantum instrument can be characterized by
  adding an ancillary system $A'$ on, say, Alice's side and
  letting $A'$ interact with an original subsystem $A$.
  Thus, the transformation
  $$\rho_{AB}\otimes\rho_{A'} \longmapsto
                               \sum_{i} p_i \proj{i}_{A'} \otimes (\rho_i)_{AB}
                            =: \sigma_{AA'B}$$
  is implemented locally by a cptp map.
  By adding a local ancilla Alice cannot
  change the quantum mutual information between her and Bob, i.e. initially
  \begin{equation}\label{lopc1}
    I(AA':B)_{\rho_{AB}\otimes\rho_{A'}}=I(A:B)_{\rho_{AB}}
  \end{equation}
  On the other hand,
  \begin{equation}\label{lopc2}\begin{split}
    I(AA':B)_{\rho_{AB}\otimes\rho_{A'}} &\geq I(AA':B)_{\sigma_{AA'B}} \\
                &=    I(A':B)_{\sigma} + I(A:B|A')_{\sigma}             \\
                &\geq I(A:B|A')_{\sigma}                                \\
                &=    \sum_i p_i I(A:B)_{(\rho_i)},
  \end{split}\end{equation}
  where in the first line we used monotonicity of $I$ under
  local operations, in the second line we used the formal ``quantum
  conditional mutual information'', and in the third and forth
  we used standard properties of the von Neumann entropy.
  Combining eqs.~(\ref{lopc1}) and (\ref{lopc2}) we obtain
  \begin{equation*}
    I(AA':B)_{\rho_{AB}} \geq \sum_i p_i I(A:B)_{\rho_i},
  \end{equation*}

  The expression on the right is the average of the total
  correlations after the instrument. We can interpret this as the
  correlation between Alice and Bob conditional on an eavesdropper
  who monitors the classical communication between them; in this
  way the common knowledge of the classical message $i$ does not
  count as correlation between Alice and Bob.
  }
\end{rem}

\section{Bipartite entanglement and classical correlations}
\label{sec:bi-entanglement}

\subsection{Quantum correlations}
\label{subsec:bi-entg}
Now we use the same method of randomisation to define an entanglement
measure. It will be the minimum noise one has to add locally to
a state $\rho$ to make it a separable state $\sigma$.
Of course, as in section~\ref{sec:bi-total} we will adopt
an asymptotic and approximate point of view:

To the disentanglement process we associate the randomising map
$R$ as in eq.~(\ref{eq:R}).
We say that $R$ \emph{$\e$-disentangles} a state $\rho$
if there is a separable state
$\sigma=\sum_{\mu} q_{\mu}\sigma_A^{\mu} \ox \sigma_B^{\mu}$
such that
\begin{equation}
  \label{e-disentanglement}
  \bigl\| R(\rho) - \sigma \bigr\|_1 \leq \e.
\end{equation}

As in section~\ref{sec:bi-total} we can (and will) restrict ourselves
to LUR, keeping in mind that the appropriate $\gamma$ in a
catalytic scenario will easily motivate a generalization to
COLUR maps.

In the previous section there was an undercurrent message that
the minimum noise we have to add is the (minimal) entropy difference
between the state and the target class. There it was product states
achievable by LUR; here we will aim at separable states achievable
by LUR (up to $\e$-approximations).
In detail, we can prove:

\begin{proposition}
  \label{prop:bi-e:lower}
  Let $T$ be an $\e$-disentangling map for $\rho^{\ox n}$.
  Then,
  \begin{equation*}\begin{split}
    \log N &\geq H(p) \geq S_e\bigl(T,\rho^{\ox n}\bigr) \\
           &\geq \inf_{\|\sigma-R(\rho^{\ox n})\|_1\leq\e}
                     \bigl( S(\sigma) - nS(\rho) - n\e\log d - \eta(\e) \bigr),
  \end{split}\end{equation*}
  where the infimum is over all COLUR maps $R$ and separable
  states $\sigma$ with $\|\sigma-R(\rho^{\ox n})\|_1\leq\e$.
\end{proposition}
\begin{proof}
  Just as in the proof of proposition~\ref{prop:bi-total:lower},
  we introduce a purification $\psi$ of $\rho$ on
  the extended system $ZAB$;
  the randomising map acts on $A^nB^n$, resulting in the state
  $$\Omega = \bigl(\id_Z^{\ox n}\ox T \bigr)\bigl(\psi^{\ox n}\bigr).$$
  As before, by the definition of the entropy exchange,
  \begin{equation*}\begin{split}
    S_e\bigl( T,\rho^{\ox n} \bigr)
                        &= S(\Omega_{Z^nA^nB^n})                   \\
                        &\geq S(\Omega_{A^nB^n}) - S(\Omega_{Z^n}) \\
                        &=    S\bigl( T(\rho^{\ox n}) \bigr)
                                     - S\bigl( \rho^{\ox n} \bigr) \\
                        &\geq S(\sigma) - nS(\rho) - n\e\log d - \eta(\e),
  \end{split}\end{equation*}
  where in the second line we have use the triangle inequality~\cite{AL71},
  and in the third line the fact that $R$ acted only on $A^nB^n$;
  in the last line we have substituted the separable state $\sigma$
  with $\| \sigma-T(\rho^{\ox n}) \|_1\leq\e$, which exists by assumption,
  and have used the Fannes inequality.
\end{proof}

\begin{proposition}
  \label{prop:bi-e:upper}
  Let $k>0$ and $T$ be a COLUR map such that $\sigma:=T(\rho^{\ox k})$
  is separable. Then for all $\e$ and sufficently large $n$ there
  exists an $\e$-disentangling COLUR map $R$ as in eq.~(\ref{eq:R}),
  with
  $$\log N \leq n\bigl( S(\sigma)-kS(\rho)+\e \bigr).$$
\end{proposition}
\begin{proof}
  We assume the form of eq.~(\ref{eq:R}) for the map $T$.
  To begin with, we have for all $n$,
  \begin{equation}
    \label{eq:manycopy-sep}
    T^{\ox n}\bigl( \rho^{\ox kn} \bigr) = \sigma^{\ox n},
  \end{equation}
  which is separable. Our goal will be to construct a COLUR map
  with the desired properties, which approximates $T^{\ox n}$.

  To this end, we use a typical projector $\Pi_1$ of $\rho^{\ox kn}$
  and a typical projector $\Pi_2$ of $\sigma^{\ox n}$:
  for sufficiently large $n$, the
  right hand side is changed by not more than $\e$ if we sandwich the
  state between $\Pi_2$, and the left hand side is changed by not
  more than $\e$ if we replace $\rho^{\ox kn}$ by
  $\widetilde\rho := \Pi_1\rho^{\ox kn}\Pi_1$. (This has the effect
  of making $\widetilde\rho \leq \frac{1}{D_1}\Pi_1$, with
  $D_1 \geq 2^{kn\bigl( S(\rho)-\e \bigr)}$.) Hence,
  $$\widehat\sigma := \Pi_2\left( T^{\ox n}\bigl( \widetilde\rho \bigr) \right)\Pi_2$$
  satisfies $\bigl\| \widehat\sigma - \sigma^{\ox n} \bigr\|_1 \leq 2\e$.

  Since $\widehat\sigma$ is supported on a subspace of dimension
  $D_2 = \tr\Pi_2 \leq 2^{n(S\bigl( \sigma)+\e \bigr)}$, we alter it
  again only by not more than $\e$ if we restrict it to the subspace
  where it is $\geq \e/D$; denote the corresponding projector
  $\Pi_3$ and let
  $\widetilde\sigma := \Pi_3\widehat\sigma\Pi_3$.

  Now we are in a position to use the operator Chernoff bound
  once more: we understand the ensemble of unitaries defining $T^{\ox n}$,
  $$W = U_I\ox V_I=(U_{i_1}\ox\cdots\ox U_{i_n})\ox(V_{i_1}\ox\cdots\ox V_{i_n})$$
  as a random variable with probability density $p(W) = p_I = p_{i_1}\cdots p_{i_n}$.
  Now we can define random operators
  $$X := D_1 \Pi_3\Pi_2 W\bigl( \widetilde\rho \bigr)W^\dg \Pi_2\Pi_3,$$
  which by the above obey $0\leq X\leq \1$, and
  $$\EE X = D_1\widetilde\rho
          \geq \e\frac{D_1}{D_2}\Pi_3
          \geq \e 2^{-n\bigl(S(\sigma)-kS(\rho)+2\e\bigr)}\Pi_3.$$

  Hence, for independent realisations $X_1,\ldots,X_N$ of $X$,
  lemma~\ref{lemma:opChernoff} gives
  \begin{equation*}\begin{split}
    \Pr&\left\{ \frac{1}{N}\sum_{j=1}^N X_j \not\in[(1\pm\e)\EE X] \right\} \\
       &\phantom{===}
        \leq 2d^n
              \exp\left( -N \e 2^{-n\bigl(S(\sigma)-kS(\rho)+2\e\bigr)}\e^2/2 \right).
  \end{split}\end{equation*}
  Hence, for $N = 2^{n\bigl(S(\sigma)-kS(\rho)+3\e\bigr)}$
  (and sufficiently large $n$), this
  probability is less than $1$; this means that
  there are unitaries $W_1,\ldots W_N$ form the original ensemble
  of product unitaries, such that
  $$(1-\e)\widetilde\sigma
       \leq \frac{1}{N} \sum_{j=1}^N W_j \widetilde\rho W_j^\dg
    \leq (1+\e)\widetilde\sigma.$$
  This statement, however, yields
  $$\left\| \frac{1}{N} \sum_{j=1}^N W_j \rho^{\ox kn} W_j^\dg
                                          - \sigma^{\ox n} \right\|_1 \leq 4\e,$$
  and we are done.
\end{proof}

\begin{rem}
  \label{rem:S_e}
  {\rm
  By the same proof technique as in propositions~\ref{prop:bi-total:upper}
  and~\ref{prop:bi-e:upper} one can show that for many independent
  copies of a COLUR map $T$ (acting on as many copies of a state $\rho$),
  the entropy exchange has, in the asymptotic
  limit, the actual character of a classical entropy rate, in the following
  sense: the action of the map $T{\ox n}$ on a purification of $\rho^{\ox n}$
  is approximated by a different COLUR map with $N$ terms, where
  $$\log N \leq n\bigl( S_e(T,\rho) + \e \bigr).$$
  \qed
  }
\end{rem}

These two propositions can be summarized in the following theorem.
Let us define, for given state $\rho$, integer $n$ and $\e>0$,
$N(n,\e)$ as the smallest $N$ such that there exists an
$\e$-distentangling COLUR map as in eq.~(\ref{eq:R}). Then, the
\emph{entanglement erasure} of $\rho$ is defined as the minimal
asymptotic noise rate needed to turn $\rho$ into a separable
state:
$$E_{\rm er}(\rho)
   := \sup_{\e>0}\limsup_{n\rar\infty} \frac{1}{n}\log N(n,\e).$$
As usual in unformation theory, we also define the \emph{optimistic
entanglement erasure} by replacing the $\limsup$ by the $\liminf$
in the previous formula:
$$\underline{E}_{\rm er}(\rho)
   := \sup_{\e>0}\liminf_{n\rar\infty} \frac{1}{n}\log N(n,\e).$$

\begin{theorem}
  \label{thm:bi-e:partialresult}
  For all bipartite states $\rho=\rho_{AB}$,
  $$\underline{E}_{\rm er}(\rho)
       \geq \sup_{\e>0}\limsup_{n\rar\infty}
                       \inf_{\|\sigma-R(\rho^{\ox n})\|_1\leq\e}
                       \frac{1}{n}S(\sigma) - S(\rho),$$
  where the infimum is over all COLUR maps $R$ and separable states $\sigma$;
  $$E_{\rm er}(\rho)
       \leq \liminf_{n\rar\infty} \inf_{\sigma=R(\rho^{\ox n})}
                       \frac{1}{n}S(\sigma) - S(\rho),$$
  with the infimum is again over all COLUR maps $R$ and
  separable states $\sigma$.
  \qed
\end{theorem}

We conjecture (without proof, at the moment) that the two limits
on the right hand side coincide. Note that the main difference
(apart from the uses of $\liminf$ and $\limsup$) is that in the
one we consider maps taking the original state to perfectly
separable states, while in the other we still allow $\e$-approximations
(which is why we need to include the $\e$ in the formula).
If this conjecture turns out to be true
we have warranted our intuition from the beginning of this section that
the entanglement erasure is the minimal entropy one has to ``add'' to the
state to make it separable.

It remains as a major open problem to prove this conjecture,
and perhaps to find a single-copy optimisation formula for
the entanglement erasure $E_{\rm er}$.

\subsection{Classical correlations}
\label{subsec:bi-class}
Now we want to use the same approach to define
and study the \emph{classical correlation} content of a quantum
state. The intuitive idea here is that what is left of the correlations
after erasing the quantum part ought to be addressed as the classical
correlations. In particular, a separable state has no quantum
correlations, so its total correlation (quantum mutual information)
should be addressed as classical correlation.

This motivates not one, but two definitions of classical correlations.
In the one we consider separable states $\sigma$ such that
there exists an LUR map $R$ such that
\begin{quote}
  (a) \quad $\bigl\| R(\rho^{\ox n}) - \sigma \bigr\|_1 \leq \e$,
\end{quote}
in the other, any local cptp map $T=T_A\ox T_B$ with
\begin{quote}
  (b) \quad $\bigl\| T(\rho^{\ox n}) - \sigma \bigr\|_1 \leq \e$.
\end{quote}
Then let
\begin{align*}
  {C\ell}_{\rm er}(\rho)
          &:= \sup_{\e>0}\limsup_{n\rar\infty}\sup_{\sigma\text{ s.t. (a)}}
                                                     \frac{1}{n}I(A:B)_\sigma \\
  {C\ell}^*_{\rm er}(\rho)
          &:= \sup_{\e>0}\limsup_{n\rar\infty}\sup_{\sigma\text{ s.t. (b)}}
                                                     \frac{1}{n}I(A:B)_\sigma.
\end{align*}
In words, ${C\ell}_{\rm er}(\rho)$ is the largest asymptotic
total erasure cost of (near-)separable states accessible from many
copies of $\rho$ by LUR, while ${C\ell}^*_{\rm er}(\rho)$
extends the maximisation over all states accessible by arbitrary
local operations (but, as in LUR, no communication or correlation).

Of course, we use the quantum mutual information to measure the
total correlations of the resulting near-separable state, because
of theorem~\ref{thm:bi-total}. There are also ``optimistic'' versions
of these definitions, denoted $\underline{C\ell}_{\rm er}$ and
$\underline{C\ell}^*_{\rm er}$, by replacing the $\limsup$
by $\liminf$; but here we will not talk about these variants.

\subsection{The pure state case}
\label{subsec:pure}
For a bipartite pure state, $\psi=\proj{\psi}$,
$\ket{\psi}=\sum_{i} \sqrt{\lambda_i}\ket{i}\ket{i}$
in Schmidt form, the total correlation is
$I(A:B)=2S(\psi_A)=2E(\psi)=2H(\lambda)$ (with $\psi_A=\tr_B\psi$),
i.e., twice the entropy of entanglement. We will show that both
the quantum and the classical correlations are equal to $E(\psi)=H(\lambda)$,
the entropy of entanglement. This is to be expected in the light of
our introductory example and from the fact of entanglement
concentration~\cite{BBPS}: indeed, for many copies of $\psi$, both
Alice and Bob can, without much distortion of the state, restrict to
their respective typical subspaces, and share a state which is pretty
much maximally entangled, at which point the reasoning of the
introduction should hold.

In rigorous detail, both Alice and Bob have typical subspace projectors
$\Pi_A$ and $\Pi_B$ for their reduced states $\psi_A^{\ox n}$
and $\psi_B^{\ox n}$, respectively, according to
lemma~\ref{lemma:typical} in the appendix. Because of that result,
we have that $\tr\bigl( \psi^{\ox n}\Pi_A\ox\Pi_B \bigr) \geq 1-\e$
for large enough $n$, and the state
$\ket{\Phi} := \Pi_A\ox\Pi_B\ket{\psi}^{\ox n}$ has Schmidt-rank
$D\leq 2^{n(S(\psi_A)+\e)}$. On the other hand, by the gentle
measurement lemma~\ref{lemma:gentle},
$\bigl\| \Phi - \psi^{\ox n} \bigr\|_1 \leq \sqrt{8\e}=:\d$.

Now a pure state of Schmidt-rank $D$ can always be disentangled
by a local phase randomisation using $D$ equiprobable unitaries:
if $\ket{\Phi}=\sum_{j} \sqrt{f_j}\ket{j}_A\ket{j}_B$, $\Phi=\proj{\Phi}$,
we let $U_k:=\sum_{j} e^{2\pi ijk/D} \proj{j}$, and have
$$\frac{1}{D}\sum_{k=1}^D (U_k\ox\1)\Phi(U_k\ox\1)^\dg
                   = \sum_{j} f_j \proj{j}_A\ox\proj{j}_B.$$
Hence, applying this same randomisation map to $\psi^{\ox n}$
will $\d$-disentangle this state.

On the other hand, let an $\e$-disentangling map $R$ for
$\psi^{\ox n}$ be given. Then, just as in the proof
of proposition~\ref{prop:bi-total:lower},
\begin{equation*}\begin{split}
  \log N &\geq H(p) \geq S_e\bigl( R,\psi^{\ox n} \bigr) \\
         &\geq S\bigl( R(\psi^{\ox n}) \bigr) - S(\psi^{\ox n}) \\
         &\geq S(\sigma) - n\e\log d - \eta(\e) - 0 \\
         &\geq S(\sigma_A) - n\e\log d - \eta(\e) \\
         &\geq S\bigl( R(\psi^{\ox n})_A \bigr) - 2n\e\log d - 2\eta(\e) \\
         &\geq S(\psi_A^{\ox n}) - 2n\e\log d - 2\eta(\e) \\
         &\geq n\bigl( S(\psi_A) - 2\e\log d - 2\eta(\e) \bigr),
\end{split}\end{equation*}
using, in this order: the triangle inequality in the second line,
then the Fannes inequality (with the separable state $\sigma$ which
we assume to exist $\e$-close to $R(\psi^{\ox n})$), then
the inequality $S(\sigma_{AB})\geq S(\sigma_A)$ for separable
states (this is implied by the majorisation result
of~\cite{nielsen:kempe}), then the Fannes inequality once more and finally
the fact that the local entropy can only increase since we use
a locally unital map.

Letting $n\rar\infty$ and $\e\rar 0$, these considerations prove that
$E_{\rm er}(\psi) = \underline{E}_{\rm er}(\psi) = E(\psi) = S(\psi_A)$.
\par\medskip
By a similarly simple consideration, we can also
calculate the classical correlation of $\psi$ (up to one
only conjectured entropic inequality):

First, by simply locally dephasing the state $\psi^{\ox n}$ in its Schmidt
basis, we can obtain a separable, perfectly correlated state $\sigma^{\ox n}$,
which has as its quantum mutual information
$$I(A:B)_{\sigma^{\ox n}} = S(\psi_A^{\ox n}) = nE(\psi).$$
On the other hand, to show that this is (asymptotically)
optimal, we need to consider local operations $T_A$ and $T_B$
(now completely general, in the spirit of the definition
of ${C\ell}^*_{\rm er}$), such that $\tau = (T_A\ox T_B)(\psi^{\ox n})$
is close to a separable state $\sigma$: $\| \tau-\sigma \|_1\leq \e$.

By implementing the local operations as local unitaries $U_A$,
$U_B$, with ancillas which we keep for reference (compare figure
1), we preserve the purity of the overall state: the output state
$\ket{\vartheta}=(U_A\ox U_B)(\ket{0}_a\ket{\psi}\ket{0}_b)$ is
the purification of $\tau$. Hence (by Uhlmann's theorem) there is
a purification $\ket{\zeta}$ of $\sigma$ such that $\|
\vartheta-\zeta \|_1 \leq \e'$, with a $\e'$ universally dependent
on $\e$ \cite{Uhlmann}. Now, invoking the Fannes inequality a
couple of times,
\begin{equation*}\begin{split}
  nE(\psi) &=    E(\vartheta) \\
           &\geq E(\zeta) - n\e'\log d - \eta(\e') \\
           &\stackrel{*}{\geq}
                 I(A_1:B_1)_\sigma - n\e'\log d - \eta(\e') \\
           &\geq I(A_1:B_1)_\tau - n(3\e+\e')\log d - 3\eta(\e) - \eta(\e').
\end{split}\end{equation*}
Division by $n$, and letting $n\rar\infty$ (such that $\e,\e'\rar 0$),
yields the claim that the mutual information rate can asymptotically not
exceed $E(\psi)$. The inequality marked $*$ in the third line we were
not able to prove rigorously (it is easily seen to be true in
a great number of cases) --- it is codified in the following
conjecture, which we think is very plausible.
\begin{conjecture}
  \label{conj:inequality}
  For pure entangled state $\psi=\psi_{AB}$, and local operations
  $T_A$, $T_B$, such that $\sigma=(T_A\ox T_B)(\psi)$ is separable,
  $$I(A:B)_{\sigma} \leq E(\psi).$$
  \begin{figure}[ht]
    \includegraphics[width=5cm]{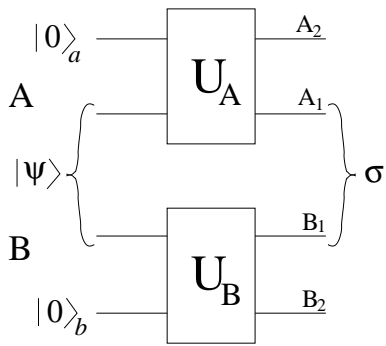}
    \caption{One can locally implement the cptp maps $T_A$ and $T_B$
             using ancillas and unitaries. These unitaries rotate the
             initial pure state $\ket{\psi}$ to a pure
             state $\ket{\zeta}=(U_A\ox U_B)(\ket{0}_a\ket{\psi}\ket{0}_b)$,
             which hence has the same entanglement as $\ket{\psi}$.
             The conjecture is thus a statement about the pure state
             $\zeta$: relative to $\zeta$, it states that
             $I(A_1:B_1) \leq S(A_1A_2)$.}
    \label{fig:conjecture}
  \end{figure}
\end{conjecture}

The major difficulty of proof stems from the fact, that Alice and
Bob may use quite general local operations if their goal is to
maximise the classical correlation, e.g. they may apply local unitaries
involving ancillas, i.e. enlarge Hilbert space (see figure 1).
If they don't do this, let's say for example that Alice acts only on her
typical subspace: then she cannot increase her local entropy
above $n\bigl( S(\psi_A)+\e \bigr)$, which also is an upper bound
for the mutual information of the separable state. In general, of course,
we would like to be able to avoid such an assumption, and indeed the
feeling is that going out of the typical subspace is suboptimal
anyway.

\section{General properties of quantum and classical correlation;\protect\\
         comparison with other entanglement measures}
\label{sec:comparison}

\subsection{Total correlation}
About the total correlation $C_{\rm er}(\rho) = I(A:B)_\rho$ of a state
we know most, primarily so because we have a usable formula.
For example, because of strong subadditivity, it is monotonic under
local operations, and in remark~\ref{rem:total:LOPC-mono} we have
already argued that monotonicity extends to local operations
and public communication.

Again because of its coinciding with the quantum mutual information,
we can easily relate the total correlation to distillability
measures of quantum states, namely total distillable
correlation, distillable secret key and distillable entanglement
(which are decreasing in this order):
$$I(A:B) \geq {\rm CR}(\rho) \geq K(\rho) \geq E_D(\rho).$$
(for the second quantity, the common randomness ${\rm CR}(\rho)$
in a state, see~\cite{igorandme1};
for the third and fourth, the distillable key $K(\rho)$ and
the distillable entanglement $E_D(\rho)$, see the recent
results in~\cite{igorandme2}).

\subsection{Quantum \&\ classical correlations}
Our theorem~\ref{thm:bi-e:partialresult} narrows down the entanglement
erasure up to the regularisation and getting rid of $\e$. This is not
good enough to decide any of the properties we would like an
entanglement measure to have --- in the first place, monotonicity
under local operations and classical communication. Similarly,
we don't know how to prove or disprove convexity of $E_{\rm er}$
(a situation much in contrast to the total correlations).

On the other hand, these properties are easily seen for the
second variant of our classical correlation quantity,
${C\ell}^*_{\rm er}$: it is monotonic under
local operations (no communication allowed, of course),
and it is convex.
\par\medskip
Once more, we have at present little to offer in terms of comparing
the erasure (quantum and classical) correlation measures to
other quantifications of entanglement and classical correlation;
clearly, we would like $E_{\rm er}$ to be an upper bound on
the distillable entanglement, and some version of the classical
correlation to be an upper bound on the distillable secret key.
It has been suggested~\cite{VPRK,Gdansk-massive-paper} that the
(regularised) relative entropy of entanglement should relate to
the entanglement erasure --- while this would be a most interesting
result, we see no clear evidence either way.
\par\medskip
An interesting question arises when we return to the pure state
example of the introduction, where the total correlations
could be erased neatly in two steps: first by adding the minimal
noise to dephase the state, and then going on from there
adding noise to classically decorrelate it. We have seen that
for pure states this is so generally, even for the asymptotic cost.
But a priori, the definitions of quantum correlations $E_{\rm er}$ and
classical correlations ${C\ell}_{\rm er}$ require us to target
quite different separable states; figure 2
illustrates this point.

\begin{figure}[ht]
  \includegraphics[width=6cm]{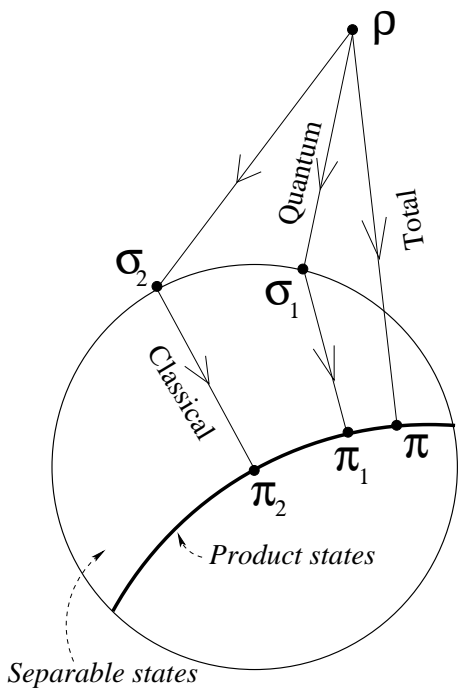}
  \caption{Starting from $\rho$, this figure illustrates the different
           objectives one has when considering (i) the total correlations,
           (ii) the quantum correlations, and (iii) the classical correlations.
           For this purpose we have ignored the subtleties of the asymptotics,
           and symbolise the noise required to go from one point in state space
           to another by their distance.
           Then for (i) we seek the shortest way (minimal noise)
           from $\rho$ to the manifold of
           product states (and we expect the target $\pi$ to be
           $\approx \rho_A\ox\rho_B$); for (ii) we seek instead the
           shortest way from $\rho$ to the convex set of separable states,
           and going to the optimal point $\sigma_1$ and from there
           on to a product state $\pi_1$
           may in total yield a suboptimal erasure procedure.
           Finally, for (iii), we want to go from $\rho$ to a separable state
           $\sigma_2$ of maximal correlation (=distance from product states).
           Even if the transition from $\rho$ to $\sigma_2$ is done by a local
           randomising map, it could be that the noise cost is significantly
           larger than that of going from $\rho$ to $\sigma_1$.
           \protect\\
           For pure state $\rho=\psi$ we have argued in
           subsection~\ref{subsec:pure}, that all three optimal paths
           coincide, and that in fact
           $E_{\rm er}(\psi) = {C\ell}^*_{\rm er}(\psi)
              = \frac{1}{2}C_{\rm er}(\psi) = E(\psi)$.}
  \label{fig:figure}
\end{figure}

Heuristically:
for an initial state $\rho$, we could have a strategy of adding
local noise to turn it into a separable state $\sigma$ (or close to).
Theorem~\ref{thm:bi-e:partialresult} indicates that the cost
will asymptotically be the entropic gap between
$\sigma$ and $\rho$: $S(\sigma)-S(\rho)$. In the spirit of
the introductory example, we then got further and completely
decorrelate $\sigma$; according to theorem~\ref{thm:bi-total}
this will cost $I(A:B)_{\sigma}=S(\sigma_A)+S(\sigma_B)-S(\sigma)$
bits of noise. Hence the total cost of this two-step process will be
$$S(\sigma_A)+S(\sigma_B)-S(\rho),$$
whereas if we had destroyed the correlations in one go, we would have
spent noise amounting to
$$I(A:B)_\rho = S(\rho_A)+S(\rho_B)-S(\rho),$$
which is in general smaller.
We have equality if the optimal disentangling map does not increase
the local entropies (or, in the asymptotic picture, only by a sublinear amount).
While this seems reasonable to expect, we have no argument in favour
of this expectation.
\par\medskip
Finally, is it true that the quantum correlation, measured
by the entanglement erasure $E_{\rm er}$, is always smaller
or equal to the classical correlation? Our and perhaps the reader's
intuition would answer yes, but to prove this from our definitions
seems not obvious.

\section{Multipartite correlations}
\label{sec:multiparty}
By obvious generalisations of the approaches presented in the previous
two sections one can also easily define total correlation and entanglement
measures for more than two parties in the many-copy limit.

We don't want to go into too much detail here but discuss an aspect of
the total correlation measure $C_{\rm er}(\rho)$ of a state
$\rho_{A_1\ldots A_p}$ of $p$ parties:

By easy generalisations of propositions~\ref{prop:bi-total:lower}
and~\ref{prop:bi-total:upper} (and remark~\ref{rem:LUN}), one
obtains that
\begin{equation}
  \label{eq:total}
  C_{\rm er}(\rho) = \sum_{i=1}^p S(A_i) - S(A_1\ldots A_p).
\end{equation}
As before, this asymptotic measure does not not depend
on the details of definition, and we find a generalisation of
the fact that the randomisation can be performed by one party
alone in the bipartite case: the parties can decorrelate
themselves locally one by one from the rest, and the individual
costs add up to $C_{\rm er}$ of eq.~(\ref{eq:total}).
In detail: let $A_1$ decorrelate herself from $A_2\ldots A_p$
using $I(A_1:A_2\ldots A_p)$ bits of randomness (by
theorem~\ref{thm:bi-total}); then let $A_2$ decorrelate himself
from $A_3\ldots A_p$ using $I(A_2:A_3\ldots A_p)$; etc.
Then adding up these quantities yields obviously
the right hand side of eq.~(\ref{eq:total}).

\section{Discussion}
\label{sec:discussion}
In this paper we have addressed the problem of an operational
definition of the total, quantum and classical amount of
correlation in a bipartite quantum state. We have shown that the
above quantities can be defined via the amount of noise that is
required to destroy the correlations.

We have proved that the total
correlation in a bipartite quantum state, measured by the
asymptotically minimal amount of noise needed to erase the
correlation, equals the quantum mutual information $I(A:B)$. Thus,
our approach gives the first clear operational definition of
$I(A:B)$ for any given state. This even lead to an operational proof
of strong subadditivity; it is an interesting question whether 
the equality conditions derived recently~\cite{HJPW04} can be derives
in this way, too.

Then we extended our approach
to definitions of the quantum (entanglement) and classical correlation
content: after definitions of these quantities in the spirit of
erasure, by the noise needed to destroy the entanglement, and the
maximum correlation left after destroying the entanglement,
we proved partial results on these quantities, and
related them to other entanglement and correlation measures.
In that context, we also put forward the conjecture
that the amount of quantum correlations is always at most
as large as the amount of classical correlations.
For pure states we have verified, up to a plausible conjectured
information inequality for separable states,
that the proposed quantum and classical correlation measures
coincide with the entropy of entanglement. In general, we had
to leave open the questions of LO(CC) monotonicity and
convexity of $E_{\rm er}$ and ${C\ell}_{\rm er}$.
(That ${C\ell}^*_{\rm er}$ is monotonic under local operations
is, however, trivial from the definition!)
\par\medskip
The reader who is acquainted with the work of the Horodecci, Oppenheim
and Sen~\cite{H3O} will sense that there is a relation between their
``thermodynamical'' approach to correlations via extractable work
(=purity), and ours, even though superficially we seem to go in opposite
directions: we consider the entropy increase necessary to destroy
correlations --- and this directly gives a correlation measure;
in the approach of~\cite{H3O} the purity content decreases
as one restricts the set of allowed operations, and the ``total
correlation'' appears as a \emph{deficit} between global operations
and local operations. If one allows also communication,
the deficit is a quantum correlations measure.
Recently, however, these authors have been able to relate
this latter deficit to the entropy production when turning
the given state into a product via certain LOCC
maps~\cite{Gdansk-massive-paper}. Via Landauer erasure, this
now looks a lot more like our model, and inded it seems to be the case that
by their including classical communication, \cite{Gdansk-massive-paper}
allows for a wider class of destructive operations, and
consequently, the resulting entanglement
measure is no larger than our $E_{\rm er}$. This makes the lower
bound from \cite{Gdansk-massive-paper} applicable, yielding that
the entanglement erasure is lower bounded by the relaive
entropy of entanglement (with respect to the separable set).
It remains to be investigated whether there is indeed a gap between them
or whether the difference is washed out in the asymptotic
limit involved in both definitions.

\acknowledgments
The authors acknowledge support from the U.~K.
Engineering and Physical Sciences Research Council
(IRC ``Quantum Information Processing'') and the EU under European Commission
project RESQ (contract IST-2001-37559). SP and AW, and also BG,
gratefully acknowledge the hospitality of the Isaac Newton Institute
for Mathematical Sciences, during the programme
``Quantum Information Science'' (16/08-17/12 2004), where part
of the present work was done.

Thanks to C. H. Bennett, I. Devetak and N. Linden for
interesting conversations on the subject of this paper,
and special thanks to the Horodecki family and J. Oppenheim for
making available their paper~\cite{Gdansk-massive-paper}
prior to publication and for discussions comparing their and
our approach.

\appendix

\section{Typicality.\protect\\ Operator Chernoff bound.\protect\\
         Miscellaneous results}
\label{app:operators}
From~\cite{Schumacher} we cite
the following definitions and properties of
\emph{typical subspaces}:

For the state density operator $\rho$ choose a diagonalisation
$\rho = \sum_{i} p_i \proj{i}$ (such that $S(\rho)=H(p)$).
Then, with $I=i_1\ldots i_n$ and
\begin{align*}
  p_I      &= p_{i_1}\cdots p_{i_n},           \\
  \proj{I} &= \proj{i_1}\ox\cdots\ox\proj{i_n},
\end{align*}
$\rho^{\ox n} = \sum_{I} p_I\proj{I}$. We call (with
$\e>0$ fixed implicitly) a state $\ket{I}$ \emph{typical}, if
$$\bigl| -\log p_I - n S(\rho) \bigr| < \e n.$$

We define the {\it $\epsilon$-typical subspace}
to be the subspace spanned by all typical states, and
$\Pi$ to be the orthogonal projector onto the typical
subspace ($n$ and $\e$ as before implicit).

The following theorem states the properties of the
typical subspace and its projector $\Pi$ (which can easily
be proved by the definitions and the law of large numbers):

\begin{lemma}[Typical subspace theorem]
  \label{lemma:typical}
  For any state $\rho$, integer $n$ and $\e>0$ let $\Pi$ the typical
  subspace projector. Then:
  \begin{itemize}
  \item For all $\d>0$ and sufficiently large $n$,
    $$\tr\bigl( \rho^{\ox n}\Pi \bigr) \geq 1-\d.$$
    In other words, by enlarging $n$ the probability of $\rho$ to
    be found in the typical subspace can be made as close to $1$
    as desired.

  \item For sufficiently large $n$, the dimension of the typical
    subspace equals $\tr\Pi$, and satisfies
    $$2^{n(S(\rho)-\epsilon)}\leq \tr\Pi \leq 2^{n(S(\rho)+\epsilon)}.$$
    Indeed, for all $n$,
    $$2^{n(S(\rho)-\epsilon)}\Pi
             \leq \Pi\rho^{\ox n}\Pi
                       \leq 2^{n(S(\rho)+\epsilon)}\Pi.$$
  \end{itemize}
\end{lemma}

\begin{lemma}[Gentle measurement~\cite{winter:qstrong}]
  \label{lemma:gentle}
  Let $\rho$ a density operator with $\tr\rho\leq 1$, and $X$ an operator
  with $0\leq X\leq \1$, such that $\tr\rho X \geq \tr\rho-\d$, then
  $$\bigl\| \rho - \sqrt{X}\rho\sqrt{X} \bigr\|_1 \leq \sqrt{8\d}.$$
  (The factor $8$ can be improved to $4$: see~\cite{ogawa:nagaoka}.)
Here the operator order is defined by saying that $X \geq Y$ iff
$X-Y$ is positive semidefinite. This is a partial order. The
interval $[ A;B ]$ is defined as the set of all operators $X$ such
that $A \leq X$ and $X \leq B$.
\end{lemma}

Furthermore, we shall make use of the following result:
\begin{lemma}[Operator Chernoff bound~\cite{AW02}]
  \label{lemma:opChernoff}
  Let $X_1,\ldots X_N$ be i.i.d.~random variables taking values in the
  operator interval $[0;\1]\subset {\cal B}(\CC^d)$ and with
  expectation $\EE X_i = M \geq \mu \1$. Then, for $0\leq \e\leq 1$,
  and denoting $\overline{X}=\frac{1}{N}\sum_{i=1}^N X_i$,
  \begin{align*}
    \Pr\bigl\{ \overline{X} \not\leq (1+\e)M \bigr\}
          &\leq d\, \exp\left( -N \frac{\mu\e^2}{2} \right), \\
    \Pr\bigl\{ \overline{X} \not\geq (1-\e)M \bigr\}
          &\leq d\, \exp\left( -N \frac{\mu\e^2}{2} \right).
  \end{align*}
\end{lemma}


\begin{thebibliography}{00}
\bibitem{CMI} Quantum
    mutual information is a straightforward formal generalization of a
    classical concept, which has clear operational definition:
    it quantifies the rate of information transmission via
    classical noisy channel.
\bibitem{Uhlmann} Uhlmann's theorem is usually formulated in
    terms of fidelity $F$. Here we use the fact that fidelity and
    trace norm are two equivalent distance measures for density operators,
    obeying the following relations:
    $F(\theta,\zeta)^2=1-\|\theta - \zeta\|_1^2$
    for pure states and
    $\|\sigma - \tau\|_1 \geq 1-F(\sigma,\tau)$ for mixed states.
\end{thebibliography}
\end{document}